\documentclass{PoS}

\title{Search for exclusive photoproduction of $Z_c(3900)$ at COMPASS}

\ShortTitle{Search for exclusive photoproduction of $Z_c(3900)$ at COMPASS}

\author{\speaker{Alexey Guskov}%
        \thanks{on behalf of the COMPASS collaboration.}\\
       Joint Inst. for Nuclear Research (RU)\\
       E-mail: \email{alexey.guskov@cern.ch}}

\abstract{The $Z_{c}(3900)$ hadron state has been found by the BES-III and Belle experiments in the decay
of the hadron state with higher mass.
The first attempt to search for the direct exclusive production of the $Z_c^{\pm}(3900)$ hadron by
virtual photons has been performed in the reaction $\mu^+ N \rightarrow \mu^+ N' Z_c(3900)^{\pm} \rightarrow \mu^+ N' J/\psi \pi^{\pm}$ at COMPASS \cite{Adolph:2014hba}.
The data cover the range from 7~GeV to 19~GeV in the centre-of-mass energy
of the photon-nucleon system. The full COMPASS data set collected
with a muon beam between 2002 and 2011 has been used.  An upper limit for the
ratio $BR(Z_c^{\pm}(3900)\rightarrow J/\psi \pi^{\pm} )\times (\sigma_{ \gamma~N
\rightarrow Z_c^{\pm}(3900)~ N} /\sigma_{ \gamma~N \rightarrow J/\psi~ N})$ of $3.7\times10^{-3}$ has
been established at the confidence level of 90\%.}

\FullConference{The European Physical Society Conference on High Energy Physics\\
		22--29 July 2015\\
		Vienna, Austria}

\begin{document}
\section{Introduction}
Untill the beginning of 2000s we knew that hadron matter existing in our universe is composed of 3 quarks (baryons) or quark-anti-quark pairs (mesons). However, QCD doesn't  prohibit forms of matter such as multi-quark states, hybrid states, glueballs and so on. In the past decade a number of such "exotic" candidates, named X,Y, Z states, have been discovered in the charmonium and bottomonium sectors. The $Z_c(3900)^{\pm}$ state observed by the BES-III \cite{BES3} and Belle \cite{Belle} collaborations in 2013 is a good candidate for a tetraquark state although other explanations like a molecular state, a cusp effect, an initial-single-pion-emission mechanism were also proposed. Since all observations of the $Z$ states have been done only in decays of other hadronic states with higher masses, looking for direct production of $Z_c(3900)^{\pm}$ looks especially interesting to clarify its nature. The COMPASS experiment performed the attempt to observe exclusive photoproduction of $Z_c(3900)^{\pm}$ using the charge-exchange reaction (see Fig. \ref{fig:diag})
\begin{equation}
\label{eq1}
\mu^+ N \rightarrow \mu^+ N' Z_c(3900)^{\pm} \rightarrow \mu^+ N' J/\psi \pi^{\pm} \rightarrow \mu^+\mu^+\mu^- \pi^{\pm} N'.
\end{equation}
\section{$Z_c(3900)^{\pm}$ production at COMPASS}
The COMPASS experiment \cite{Abbon:2007pq} is situated at the M2 beam line of the CERN Super Proton Synchrotron. 
The experimental data obtained for positive muons scattering
             of 160~$GeV/c$ (2002-2010) or 200~$GeV/c$ momentum (2011) off solid $^6$LiD
             (2002-2004) or NH$_3$ targets (2006-2011) were used for looking for exclusive $Z_c(3900)^{\pm}$ production. 
 \begin{figure}
 \begin{minipage}{18pc}
   \includegraphics[width=200px]{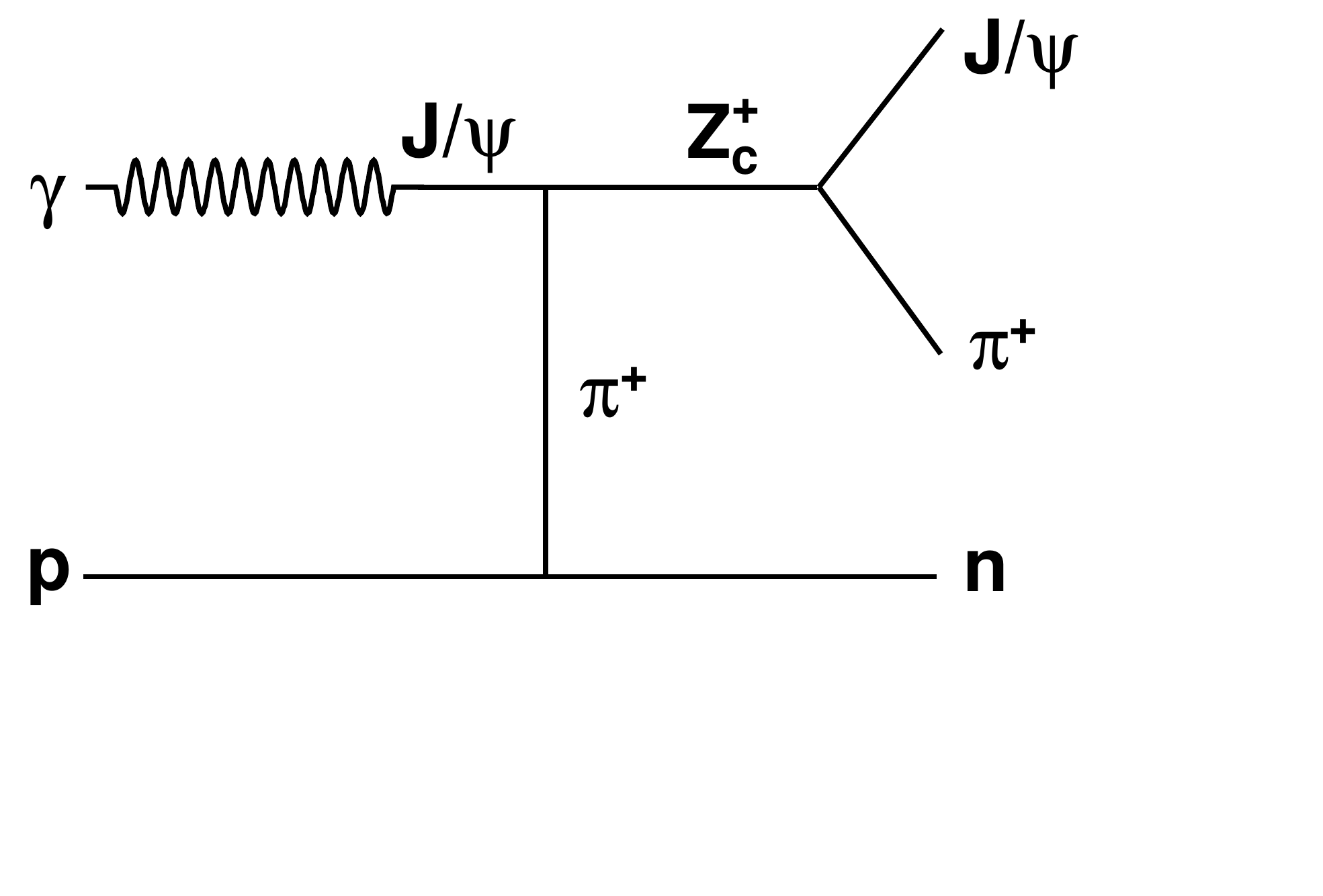}  
     \caption{\label{fig:diag}
Diagram for $Z_c^+(3900)$ production via virtual $\pi^+$ exchange.}
\end{minipage}\hspace{2pc}%
\begin{minipage}{18pc}
   \includegraphics[width=200px]{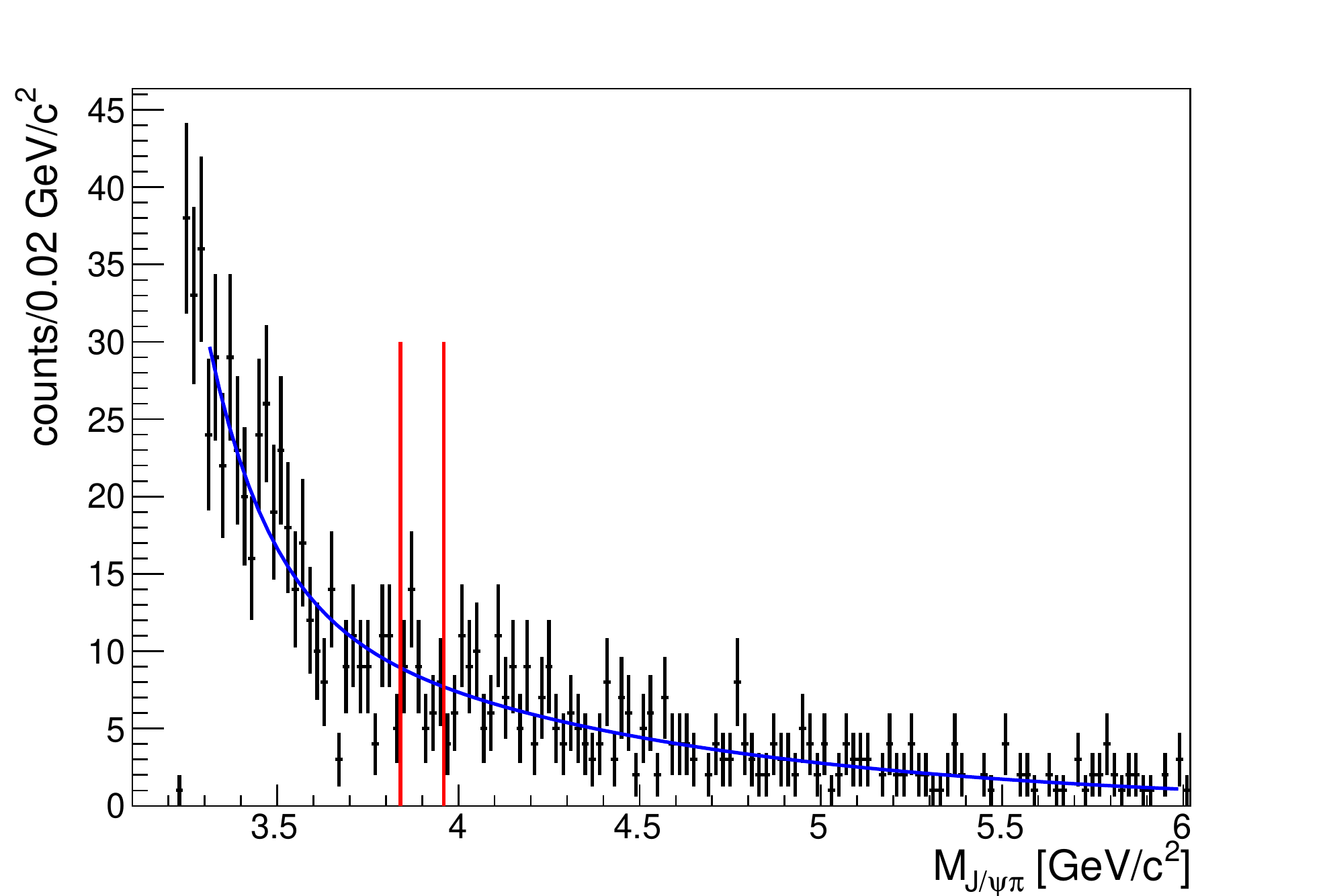}  
     \caption{\label{spectr}Mass spectrum of the $J\psi\pi^{\pm}$ \cite{Adolph:2014hba}. The searching range is shown by the vertical lines while the curve represents the background fitting.}
\end{minipage}
 \end{figure}
The mass spectrum for $J/\psi \pi^{\pm}$ events is shown in Fig.~\ref{spectr}. It does not exhibit any statistically significant resonant structure around 3.9~GeV/$c^2$. The upper limit  $N^{UL}_{Z_c}$ for the number of produced $Z^{\pm}_c(3900)$ events in the range $3.84~GeV/c^2<M_{J/\psi \pi^+}<3.96~GeV/c^2$  corresponding to a confidence level of $CL=90\%$ is estimated to be $N^{UL}_{Z_c}=15.1$ events. For the absolute normalisation of the $Z_c^{\pm}(3900)$ production rate the number of exclusively produced $J/\psi$ mesons from incoherent exclusive production 
\begin{equation}
\label{reaction2}
\gamma~N \rightarrow J/\psi~N,
\end{equation}
in the same data sample is used. The obtained result is
\begin{equation}
\label{result}
\frac{BR(Z_c^{\pm}(3900)\rightarrow J/\psi \pi^{\pm} )\times \sigma_{ \gamma~N \rightarrow Z_c^{\pm}(3900)~ N}}{\sigma_{ \gamma~N \rightarrow J/\psi~ N} }\Big{|}_{\sqrt{s_{\gamma N}}\rangle=13.8~GeV} < 3.7\times10^{-3}.
\end{equation} 
Taking into account the known cross section of the reaction \ref{reaction2}:
\begin{equation}
\label{result2}
BR(Z_c^{\pm}(3900)\rightarrow J/\psi \pi^{\pm} )\times \sigma_{ \gamma~N \rightarrow Z_c^{\pm}(3900)~ N}\Big{|}_{\langle\sqrt{s_{\gamma N}}\rangle=13.8~GeV} < 52~\mathrm{pb}.
\end{equation}

\section{COMPASS: new possibilities}
Forthcoming upgrade of the COMPASS setup related with the planned data taking  within the framework of the GPD program \cite{COMPASS_proposal} could provide new opportunities for searching for direct production of exotic charmonium-like states including $Z_c(3900)$. The system of electromagnetic calorimeters ECAL1 and ECAL2 will be extended by installation of the new large-aperture calorimeter ECAL0. Tightness of the most forward calorimeter ECAL2 will be significantly improved by reducing size of the beam hole. With the new calorimetry system we can expect much better selection of exclusive events. Searching for production of the neutral $Z_c(3900)^0$, discovered by BES-III, decaying into $J/\psi\pi^0$ will also be possible. The final states with the $\chi_c$ -mesons could also be studied.

The update will change the configuration of the target region: a 2.5 m long liquid hydrogen target surrounded by a 4 m long recoil proton detector will be used. Absence of the neutrons in the target will allow to use only the $J/\psi\pi^+$ final state for searching for $Z^{+}_c(3900)$ while $J/\psi\pi^-$ mass spectrum could be used for studies of systematic effects. The recoil proton detector CAMERA consists of two barrels of 24 scintillator slabs. This detector serves the double purpose: to reconstruct and identify recoil protons via time- of-flight and energy loss measurements. Since the reaction of exclusive production of $Z^{+}_c(3900)$ off the proton doesn't have a recoil proton in the final state, any activity in the CAMERA could serve as a veto in the offline analysis.

\section{Summary}
Search for the exclusive photoproduction of the $Z_c^{\pm}(3900)$ state using the reaction (\ref{eq1}) has been performed at COMPASS.  No statistically significant signal was found. 
Therefore an upper limit was determined for the product of the cross section of this process 
and the relative $Z_c^{\pm}(3900)\rightarrow J/\psi \pi^{\pm}$ decay probability normalized to the 
cross section of incoherent exclusive photoproduction of $J/\psi$ mesons. This result is a significant input to clarify the nature of the $Z_c^{\pm}(3900)$ state. For instance it could be converted to the upper limit of the partial width $\Gamma_{J/\psi\pi}<2.4$ MeV basing on the production model assuming vector meson dominance (VMD) \cite{MAIN}. In case $Z_c^{\pm}(3900)$ is a real hadron state, the decay channel $Z_c^{\pm}(3900)\rightarrow J/\psi \pi^{\pm}$ can not be the dominant one.
The $J/\psi\pi$ mass spectrum measured by COMPASS  was also used for the estimation of the $Z_c^+(4200)$ photoproduction cross section \cite{XY}.


\begin{thebibliography}{99}
\bibitem{Adolph:2014hba}
  C.~Adolph {\it et al.} [COMPASS Collaboration],
  Phys.\ Lett.\ B {\bf 742} (2015) 330
  [arXiv:1407.6186 [hep-ex]].
\bibitem{BES3} M. Ablikim, et al., Phys. Rev. Lett. \textbf{110}, 252001 (2013)
\bibitem{Belle} Z. Q. Liu, et al.,  Phys. Rev. Lett. \textbf{110}, 252002 (2013).
\bibitem{Abbon:2007pq} P.~Abbon et al. (COMPASS Collaboration), NIM \textbf{A577}, 455 (2007), [arXiv:0703049~[hep-ex]].
\bibitem{COMPASS_proposal} COMPASS, SPSC-2010-014/P-340
\bibitem{MAIN} Q.-Y. Lin et al., Phys. Rev. \textbf{D88}, 114009 (2013),  [arXiv:1308.6345 [hep-ph]].
\bibitem{belle100} K. Chilikin et al. (Belle Collaboration), Phys. Rev. \textbf{D88}, 074026 (2013).
 \bibitem{XY}X. Y. Wang et. al, [arXiv:1503.02125 [hep-ph]].
  
\end{thebibliography}
\end{document}